\documentclass[sigconf]{acmart}

\AtBeginDocument{%
  \providecommand\BibTeX{{%
    \normalfont B\kern-0.5em{\scshape i\kern-0.25em b}\kern-0.8em\TeX}}}

\setcopyright{acmlicensed}
\copyrightyear{2025}
\acmYear{2025}
\acmISBN{}
\acmDOI{}

\acmConference[KDD '25]{KDD '25: SIGKDD Conference on Knowledge Discovery and Data Mining}

\usepackage{amsmath,amsfonts,bm}

\def\eqref#1{equation~\ref{#1}}

\def\1{\bm{1}}

\DeclareMathAlphabet{\mathsfit}{\encodingdefault}{\sfdefault}{m}{sl}
\SetMathAlphabet{\mathsfit}{bold}{\encodingdefault}{\sfdefault}{bx}{n}

\usepackage[most]{tcolorbox}
\usepackage{caption}
\usepackage{multirow}
\usepackage{enumitem}
\usepackage{subfig}
\usepackage{color, colortbl}
\usepackage{makecell}
\usepackage[symbol]{footmisc}

\definecolor{Gray}{gray}{0.9}
\begin{document}
\newtheorem{problem}{Problem}

\title{MOPI-HFRS: A Multi-objective Personalized Health-aware Food Recommendation System with LLM-enhanced Interpretation}

\author{Zheyuan Zhang}
\affiliation{%
 \institution{University of Notre Dame}
 \city{Notre Dame}
 \state{Indiana}
 \country{USA}}
\email{zzhang42@nd.edu}

\author{Zehong Wang}
\affiliation{%
 \institution{University of Notre Dame}
 \city{Notre Dame}
 \state{Indiana}
 \country{USA}}
\email{zwang43@nd.edu}

\author{Tianyi Ma}
\affiliation{%
 \institution{University of Notre Dame}
 \city{Notre Dame}
 \state{Indiana}
 \country{USA}}
\email{tma2@nd.edu}

\author{Varun Sameer Taneja}
\affiliation{%
 \institution{University of Notre Dame}
 \city{Notre Dame}
 \state{Indiana}
 \country{USA}}
\email{vtaneja@nd.edu}

\author{Sofia Nelson}
\affiliation{%
 \institution{University of Notre Dame}
 \city{Notre Dame}
 \state{Indiana}
 \country{USA}}
\email{snelso24@nd.edu}

\author{Nhi Ha Lan Le}
\affiliation{%
 \institution{Brandeis University}
 \city{Waltham}
 \state{MA}
 \country{USA}}
\email{nhihlle@brandeis.edu}

\author{Keerthiram Murugesan}
\affiliation{%
 \institution{IBM T.J. Watson Research Center}
 \city{Yorktown Heights}
 \state{NY}
 \country{USA}}
\email{keerthiram.murugesan@ibm.com}

\author{Mingxuan Ju}
\affiliation{%
 \institution{University of Notre Dame}
 \city{Notre Dame}
 \state{Indiana}
 \country{USA}}
\email{mju2@nd.edu}

\author{Nitesh V. Chawla}
\affiliation{%
 \institution{University of Notre Dame}
 \city{Notre Dame}
 \state{Indiana}
 \country{USA}}
\email{nchawla@nd.edu}

\author{Chuxu Zhang}
\affiliation{%
 \institution{University of Connecticut}
 \city{Storrs}
 \state{CT}
 \country{USA}}
\email{chuxu.zhang@uconn.edu}

\author{Yanfang Ye} 
    \authornote{Corresponding Author.}
\affiliation{%
 \institution{University of Notre Dame}
 \city{Notre Dame}
 \state{Indiana}
 \country{USA}}
\email{yye7@nd.edu}

\settopmatter{authorsperrow=4}

\renewcommand{\shortauthors}{Zheyuan Zhang et al.}

\begin{abstract}
The prevalence of unhealthy eating habits has become an increasingly concerning issue in the United States. However, major food recommendation platforms (e.g., Yelp) continue to prioritize users’ dietary preferences over the healthiness of their choices. Although efforts have been made to develop health-aware food recommendation systems, the personalization of such systems based on users' specific health conditions remains under-explored. In addition, few research focus on the interpretability of these systems, which hinders users from assessing the reliability of recommendations and impedes the practical deployment of these systems. In response to this gap, we first establish two large-scale \textit{personalized} health-aware food recommendation benchmarks \textit{at the first attempt}. We then develop a novel framework, \textbf{M}ulti-\textbf{O}bjective \textbf{P}ersonalized \textbf{I}nterpretable \textbf{H}ealth-aware \textbf{F}ood \textbf{R}ecommendation \textbf{S}ystem (\textbf{MOPI-HFRS}), which provides food recommendations by jointly optimizing the three objectives: user preference, personalized healthiness and nutritional diversity, along with an large language model (LLM)-enhanced reasoning module to promote healthy dietary knowledge through the interpretation of recommended results. Specifically, this holistic graph learning framework first utilizes two structure learning and a structure pooling modules to leverage both descriptive features and health data. Then it employs Pareto optimization to achieve designed multi-facet objectives. Finally, to further promote the healthy dietary knowledge and awareness, we exploit an LLM by utilizing knowledge-infusion, prompting the LLMs with knowledge obtained from the recommendation model for interpretation. Extensive experimental results based on our established benchmarks demonstrate the outstanding performance of MOPI-HFRS in providing both diverse, healthy food recommendations and reliable explanations by comparison with state-of-the-art baseline methods. Our code, built benchmarks are available at \href{https://github.com/Anonymous-Be3fb6/MOPI-HFRS/tree/main}{here}.

\end{abstract}
\maketitle

\begin{figure*}[htbp!]
	\centering
	\includegraphics[width=1\linewidth]{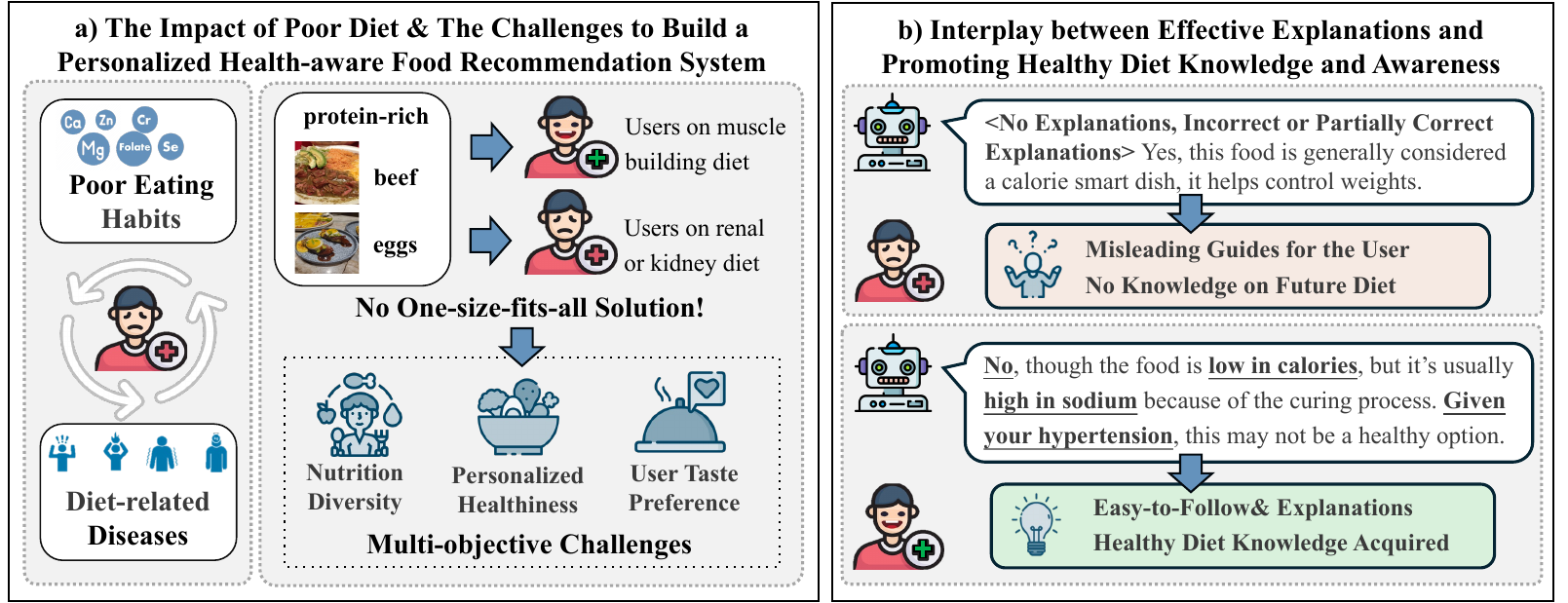}
        \vspace{-20pt}
	\caption{The illustration of the impact of a poor diet, and the comprehensive and challenging nature of the multi-objective personalized interpretable health-aware food recommendation task.}
        \vspace{-15pt}
    \label{fig:motivation}
\end{figure*}

\section{Introduction}
Diet plays a pivotal role in maintaining and enhancing human health. A balanced diet provides the essential nutrients required for optimal physical and mental well-being, while poor dietary habits can lead to numerous adverse health outcomes, highlighting the critical importance of diet in public health \cite{WHO_2021_HealthyDiet}. Despite the well-documented benefits of a balanced healthy diet, the prevalence of unhealthy eating habits remains a significant concern in modern society. Approximately 42.4\% of adults in the United States are classified as obese \cite{CDC_2020_Obesity}, and in 2017 alone, 11 million deaths and disability-adjusted life-years (DALYs) were attributable to dietary risk factors, such as high intake of sodium \cite{Afshin_2019_DietaryRisks, WHO_2023_Obesity}. It is vital to counteract this trend and encourage individuals to adopt healthier eating habits. Unfortunately, the field of health-aware food recommendation systems has not received as much attention as other domains, such as movie recommendation systems \cite{trattner2017food}. This disparity may be due to the comprehensive and challenging nature of the task: On one hand, there is not a one-size-fits-all solution for healthy food recommendations, and the system needs to balance multiple goals, such as user preference, personalized healthiness, and diversity, as depicted in Figure-\ref{fig:motivation}(a); On the other hand, the explainability of the system is crucial, as the goal of such systems is not only to inform users of the recommendations but also to explain why the options are healthier, thereby raising their awareness and knowledge towards healthy diets, as shown in Figure-\ref{fig:motivation}(b). So the question naturally occurs: \textit{How can we develop an effective and interpretable food recommendation system that simultaneously considers users’ dietary preferences, diversity, and personal health conditions?}

To answer this question, numerous existing studies \cite{ge2015health, shirai2021identifying, wang2021market2dish, bolz2023hummus, zhang2024greenrec} have attempted to define and measure the healthiness of foods. However, we argue these efforts are significantly limited in several ways, which are unable to fully address the complexities of our question at hand: First, to the best of our knowledge, none of the current works are truly personalized to users' specific health conditions, because of the inaccessibility of users' medical information. The definition of "healthy" varies significantly between individuals with different health profiles, such as the individuals with high body mass index (BMI) compared to those with low BMI, or individuals recovering from opioid misuse compared to those with chronic kidney diseases. Second, the healthiness of a food in current studies tend to be used as a hard-coded post-processing method, with the information rarely utilized during the training process, thus unable to capture more nuanced interplay between health factors and user preference. Third, existing approaches often neglect the importance of balancing health goals with individual tastes, which hinders the practical deployment of such systems, as this can lead to potential user dissatisfaction. Finally, relatively few studies focus on the interpretability of the models, whereas the interpretations are crucial in health eating habit promotion, as they help user understand and accept the recommendations, and empower the users to make better-informed decisions about their diet beyond the immediate use of the recommendation system in the future.

To overcome the aforementioned four research gaps, we first establish two large-scale food recommendation benchmark datasets, marking the first attempt to integrate users’ medical data for personalized health-aware food recommendations. Specifically, we leverage data from the National Health and Nutrition Examination Survey (NHANES) \cite{NHANES}, a publicly available dataset designed to evaluate the health and nutritional status of adults and children in the United States. Using this data, we construct \textit{the Health and Nutrition Recommendation Bipartite Graph} and two benchmark datasets derived from this graph to comprehensively integrate users’ health-related information and food records. This way we tackle the first research gap of unable to personalize the recommendation.

We then develop a novel framework, \textbf{M}ulti-\textbf{O}bjective \textbf{P}ersonalized \textbf{I}nterpretable \textbf{H}ealth-aware \textbf{F}ood \textbf{R}ecommendation \textbf{S}ystem (\textbf{MOPI-HFRS}), that bridges the gap between graph-based recommendation system and LLMs to train an interpretable food recommendation system that simultaneously considers three objectives: user preference, personalized healthiness, and nutritional diversity. Specifically, MOPI-HFRS is designed to address the remaining three gaps as follows: First, to tackle the issue of not involving health information in the training, we develop a health-aware graph structural learning module that dynamically incorporates both health-related and descriptive feature information during training. Second, to achieve the optimal recommendations while balancing the three objectives, we apply Pareto multi-objective learning for model optimization. Finally, to increase system's interpretability, we further integrate the generative power of LLMs that prompted with knowledge obtained from the recommendation tasks for explanations, raising public awareness and knowledge about healthy diets. Overall, the key contributions of this work are summarized as follows:

\begin{itemize}[leftmargin=*]
    \item \textbf{New Benchmarks.} In this work, we construct a large-scale Health and Nutrition Recommendation Graph and establish two benchmark datasets from the graph for health-aware food recommendation tasks. To our best knowledge, this is \textit{the first attempt} to incorporate users' medical information to personalize healthy food recommendations.
    \item \textbf{Novelty.} We develop an innovative and holistic framework named \textbf{MOPI-HFRS} to provide multi-objective interpretable \textit{personalized} health-aware food recommendations. This framework presents a novel health-aware graph structural learning module and a Pareto Optimization module. The knowledge gained from the recommendation process also guides a downstream LLM-enhanced reasoning task, which provides insightful interpretations for the recommendations.
    \item \textbf{Effectiveness.} Extensive experiments based on our established benchmarks, demonstrate that \textbf{MOPI-HFRS} outperforms existing state-of-the-arts baselines in both the multi-objective recommendation task and the downstream reasoning task. 
\end{itemize}

\section{Related Work}

\noindent\textbf{Health-aware Food Recommendation System.} Food recommendation is a well-established task with benchmarks like Recipe1M+ \cite{marin2021recipe1m+, salvador2017learning} and FoodKG \cite{haussmann2019foodkg}, and numerous studies \cite{tian2022reciperec, li2023user, shi2023recipemeta}. However, these primarily address dietary preferences rather than healthiness. With increasing awareness of diet health, recent efforts have aimed to integrate health metrics into recommendations, typically falling into three categories. First, some works focus on single metrics like calories or fat, as seen in Ge et al. \cite{ge2015health} and Shirai et al. \cite{shirai2021identifying}. However, these metrics alone seldom capture the complexity of a healthy diet. Second, simulated methods to generate health data, such as those by Wang et al. \cite{wang2021market2dish}, often fail to reflect real-life data distribution. Lastly, recent research \cite{bolz2023hummus, zhang2024greenrec} has incorporated international health standards to define comprehensive health scores. Yet, no universal solution exists, as foods considered healthy can still trigger adverse effects in some individuals \cite{yue2021overview}. The core challenge remains the lack of reliable user health data. Consequently, the task of creating a truly personalized health-aware food recommendation system remains open. 

\noindent\textbf{LLM-enhanced Interpretable Recommendation System.}
Interpretable recommendation systems are increasingly valued for enhancing user satisfaction and transparency. Traditional methods primarily focused on extracting attributes from user and item data, often using attention mechanisms \cite{dong2017learning}, or graph-based approaches that utilized graph neural networks for interpretability \cite{he2015trirank, chen2021temporal, wang2018ripplenet}. Recent advancements in large language models (LLMs) have further explored the potential for text generation, providing valuable insights into recommendation results. For example, Ma et al. \cite{ma2024xrec} develop a model-agnostic framework to integrate insights from collaborative filtering into LLMs; Li et al. \cite{li2023personalized} use prompt learning to provide user-understandable explanations; And Wang et al \cite{wang2024rdrec} use ids as prompt tokens for explanation generation. Retrieval-augmented generation \cite{lewis2020retrieval} has also emerged as a popular approach, especially when integrating LLMs with knowledge graphs \cite{hussien2024rag, mavromatis2024gnn}. These techniques for interpretable recommendation represent a promising direction for future research.

\section{Preliminaries}
In this section, we first elucidate the key concepts foundational to our study, followed by an exposition on the Health and Nutrition Recommendation Bipartite Graph, and two versions of benchmark datasets adapted from it.

\begin{definition}[Bipartite Graph for Recommendation]
    Given a bipartite graph, denoted as $\mathcal{G} = (\mathcal{U}, \mathcal{F}, \mathcal{E})$, where $\mathcal{U}$ denotes the user set, $\mathcal{F}$ denotes the food set, and $\mathcal{E}$ is the set of edges between $\mathcal{U}$ and $\mathcal{F}$. It can be denoted as the adjacency matrix $\mathcal{A} \in \mathbb{R}^{(|\mathcal{U}|+|\mathcal{F}|) \times (|\mathcal{U}|+|\mathcal{F}|)}$, where $\mathcal{A}_{uf} \neq 0$ means that user $u$ has an interaction with food $f$. Additionally, let $\mathcal{X}_{\mathcal{U}}$ and $\mathcal{X}_{\mathcal{F}}$ denote the features of users and foods, respectively.
\end{definition}

\begin{definition}[Signed Bipartite Graph]
    Given a signed bipartite network, \(\mathcal{G} = (\mathcal{U}, \mathcal{F}, \mathcal{E})\), where \(\mathcal{E} = \mathcal{E}^+ \cup \mathcal{E}^-\) is the set of edges between the two sets of nodes $\mathcal{U}$ and $\mathcal{F}$ . The sets \(\mathcal{E}^+\) and \(\mathcal{E}^-\) represent the sets of positive and negative edges (in our case, healthy or unhealthy edges) respectively, and \(\mathcal{E}^+ \cap \mathcal{E}^- = \emptyset\).
\end{definition}

Given the graph $\mathcal{G} = (\mathcal{U}, \mathcal{F}, \mathcal{E}, \mathcal{X}_{\mathcal{U}}, \mathcal{X}_{\mathcal{F}})$, our task is to design a machine learning model $\mathbf{F}_{\Theta}$ with parameters $\Theta$, optimized by objective functions $\mathcal{L}_1, \mathcal{L}_2, \ldots, \mathcal{L}_n$, to recommend foods to users. Specifically, the recommendation is supposed to take advantage of the associated user and food features $\mathcal{X}_{\mathcal{U}}$, $\mathcal{X}_{\mathcal{F}}$ and relational structure information in $\mathcal{E}$, and return a ranked list of foods for each user. To achieve the multiple objectives, and simultaneously optimize the model towards different goals, our focus of the multi-objective optimization is to approach Pareto optimality \cite{desideri2012multiple}, which in our case can be defined as follows: 

\begin{definition}[Pareto Optimality]
    A given set of parameters $\theta^*$ optimized by objective functions $\mathcal{L}_1, \mathcal{L}_2, \ldots, \mathcal{L}_n$ is Pareto optimal if and only if $\mathcal{L}_k(\mathcal{G}, \theta^*) \leq \mathcal{L}_k(\mathcal{G}, \theta)$ for every for objective function k, i.e., no individual loss can be minimized further without increasing at least one other loss.
\end{definition}

\begin{figure}[!t]
  \centering
  \includegraphics[width=\linewidth]{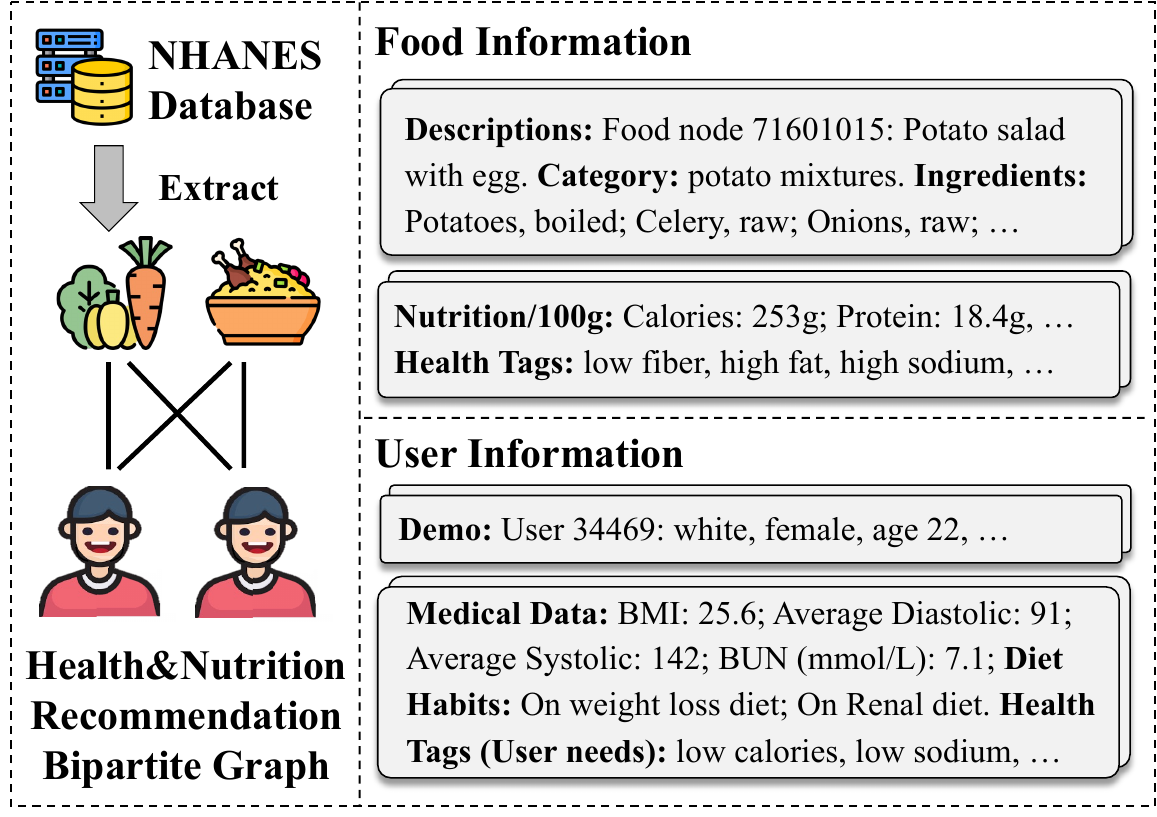}
  \vspace{-20pt}
  \caption{The schema of Health and Nutrition Recommendation Bipartite Graph.}
  \vspace{-15pt}
  \label{fig:graph}
\end{figure}

\noindent{\textbf{Constructing the Benchmark.}} 
Utilizing NHANES data from 2003 to 2020, we construct the Health and Nutrition Recommendation Bipartite Graph, the first graph dataset aiming at providing health-aware food recommendation personalized to users' specific health conditions. This graph is constructed from the food intake data and the users' medical data ranging from lab results to recorded questionnaires. To measure the healthiness of a recommended food to a user, we utilize the tagging scheme to assign user and food health tags. Specifically, on the side of food tagging, besides the aforementioned WHO and FSA standards, we also involve the standards from EU Nutrition \& Health Claims Regulation legislation \cite{EC2006} and the Codex Alimentarius Commission (CAC)\cite{FAO1985, FAO1997} to get more fine-grained standards on common nutrients. For example, the standards define that a claim that a food is low in sodium/salt may only be made where the product contains no more than 0.12 g of sodium, or the equivalent value for salt, per 100 g \cite{EC2006}. Thus foods meet such requirements in our benchmark will be tagged as "low\_sodium". On the other hand for user tagging, we use the medical data that the NHANES dataset provides. For example, if a user has high blood pressure \cite{grillo2019sodium}, or the user is on a low-salt or Renal/Kidney diet \cite{smyth2014sodium}, it indicates the user should limit their salt intake, thus the user is tagged as "low\_sodium" (representing the user needs low sodium diet). 


In this paper, we involve health tags for 16 different nutrients focusing on various health aspects, including 7 for macro-nutrients (calories, carbohydrates, protein, saturated fat, cholesterol, sugar, and dietary fiber) and 9 for micro-nutrients (sodium, potassium, phosphorus, iron, calcium, folic acid, and vitamin C, D, and B12). As can be seen in Figure-\ref{fig:graph}, besides health information, other user features include \textit{anonymized} demographic information, while the food features comprise nutritional values, and encoded textual descriptions \cite{devlin2018bert} of the foods and their categories and ingredients. For our tasks, we develop two versions of benchmarks from the graph with one considering only the macro-nutrients and the other considering both macro and micro nutrients. This is because, many domain studies and guidelines focus on macro nutrients as they are more available across all food items as well as more impactful to metabolic health \cite{howell2017calories, who2020healthydiet, ludwig2018carbohydrate}, while the full nutrients benchmark provides more comprehensive information. Hence, the two benchmarks both play an important role in practice. Further details of the tags we used and the corresponding health index and nutritional standards are available in Appendix-B.


\begin{figure*}[htbp!]
	\centering
	\includegraphics[width=1\linewidth]{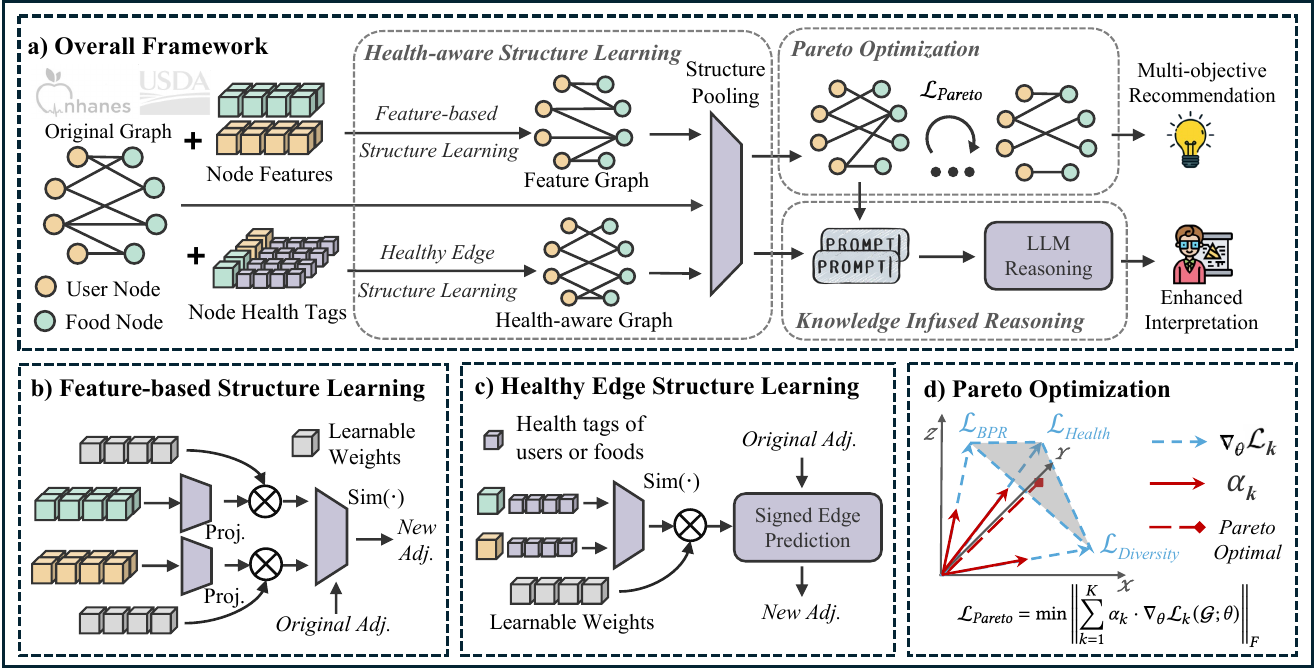}
        \vspace{-20pt}
	\caption{(a) The overall framework of MOPI-HFRS, which consists of three major components: (b) a health-aware structure learning module with feature-based structure learning, and (c) healthy edge structure learning; (d) a Pareto optimization module  and a knowledge-infused reasoning module.}
        \vspace{-15pt}
    \label{fig:framework}
\end{figure*}

\section{Methodology}
Based on our created benchmark datasets, we introduce an innovative framework for multi-objective personalized interpretable health-aware food recommendation (MOPI-HFRS), illustrated in Figure-\ref{fig:framework}. This framework incorporates three key modules: 1) a health-aware graph structure learning module, which aims to dynamically integrate descriptive, nutritional and health information to learn a refined graph structure with high data quality (detailed in Section-4.1); 2) a Pareto optimization module that balances trade-off between the multiple objectives, which provides users diversified diets that meet both their dietary preferences and specific health requirements (detailed in Section-4.2); 3) a knowledge-infused LLM-based interpretation module using the knowledge obtained from previous steps to bridge the gap between recommendations and their explanations. (detailed in Section-4.3).

\vspace{-7pt}
\subsection{Health-aware Graph Structure Learning} 

Given the properties of the real-life NHANES datasets, providing users with balanced healthy foods presents a unique challenge, requiring detailed analysis of (1) Users' diet preference and shared food nutrition patterns among users. (2) Integrating user's specific health requirements to match foods provide corresponding nutrients. (3) Reducing noise introduced during data collection. Our model addresses these three challenges by jointly leveraging the descriptive feature information (to tackle the first challenge), health information (to address the second challenge), and performing structure learning for a refined graph structure (to resolve the third challenge). As illustrated in the Health-aware Graph Structure Learning module in Figure-\ref{fig:framework}(a), the module contains three components: the feature-based graph structure learning (i.e. Figure-\ref{fig:framework}(b)), the healthy edge graph structure learning  (i.e. Figure-\ref{fig:framework}(c)), and a structure pooling layer to aggregate the two generated graphs and the origin graph structure together for downstream tasks. In the following section, we will introduce each component in details. 

\subsubsection{Feature-based Graph Structure Learning}
Users with similar backgrounds, such as ages or races, often exhibit similar dietary preferences. These preferences can serve as subtle yet informative sources to recommend foods that users or their immediate neighbors (who share the same foods with the user) haven't previously interacted with. Motivated by this observation, instead of directly using features as embedding initialization, we introduce the feature-based structure learning module. Specifically, we first employ a feature-specific projection to map the user feature and food feature $\mathbf{f}$ into the same space: 

\vspace{-5pt}
\begin{equation}
    \mathbf{f}'_i = \sigma (\mathbf{f}_i \cdot \mathbf{W}_i + \mathbf{b}_i),
\end{equation}
where \( \sigma (\cdot) \) denotes a non-linear activation function, \( \mathbf{W}_{i} \) and \( \mathbf{b}_{i} \) represent the mapping matrix and the bias vector respectively for type \( i \) (i.e., user or food). Then, we implement metric learning to obtain a learned feature similarity graph \(\mathcal{G}_{ft}\) and its adjacency matrix \(\mathcal{A}_{ft}\), defined as:

\begin{equation}
\mathcal{A}_{ft}[u,f] = \begin{cases}
\tau(\mathbf{f}'_u, \mathbf{f}'_f) & \text{if } \tau(\mathbf{f}'_u, \mathbf{f}'_f) \geq \epsilon \\
0 & \text{otherwise}
\end{cases},
\end{equation}
where \(u \in \mathcal{U}, f \in \mathcal{F}, \text{ and } \epsilon\) is the threshold that controls how similar the features are required to be included and a larger $\epsilon$ implies a more spare feature similarity graph. $\tau$ is a K-head weighted cosine similarity function defined as: 

\vspace{-5pt}
\begin{equation}
    \tau(\mathbf{f}'_u, \mathbf{f}'_f) = \frac{1}{K} \sum_{k=1}^{K} \cos (\mathbf{W}_{k} \odot \mathbf{f}'_u, \mathbf{W}_{k} \odot \mathbf{f}'_f),
\end{equation}
where \( \odot \) denotes the Hadamard product, and \( \mathbf{W}_{k} \) is a set of the learnable weights to learn the importance of different dimensions of the feature vectors. By performing the metric learning, we dynamically generate a feature graph that focuses on refine the structure to connect similar users and foods.

\subsubsection{Healthy Edge Graph Structure Learning}
While the feature-based structure learning leverages the descriptive feature similarity between users and foods, user's health information is not easily aligned with food's nutrition information, and this is where domain knowledge becomes crucial. We utilize the generated health tags (details in Section 3), and define a food healthy to a user if it contains more matching tags than converse tags (e.g. "high\_sodium" and "low\_sodium"). In this way we transform the original graph to a signed graph, where the edges are marked as positive (healthy) or negative (unhealthy). According to the balance theory \cite{cartwright1956structural}, a balanced cycle consists of an even number of negative links. In our context, the rationale is that the embeddings of the healthy foods and unhealthy foods should be aggregated differently. For example, if food A and B are both unhealthy for a user, their embeddings should be closer to each other than that of food C, assuming food C is healthy to the user. Specifically, in the aggregation phase, the embedding ${h}$ of node ${i}$ at layer ${l}$ can be defined as the concatenation of two embedding sets: The positive and negative embeddings [\(\mathbf{h}^{+(l)}_i, \mathbf{h}^{-(l)}_i\)]. When ${l}$ = 1, the positive and negative embeddings are relatively intuitive. Take the positive embeddings as an example, it can be defined as:  

\vspace{-5pt}
\begin{equation}
\mathbf{h}^{+(1)}_i = \sigma \left( \mathbf{W}^{+(1)} \left[ \sum_{j \in \mathcal{N}_i^+} \frac{\mathbf{h}^{+(0)}_j}{|\mathcal{N}_i^+|}, \mathbf{h}^{+(0)}_i \right] \right),
\end{equation}
where $\sigma$ is a non-linear activation function and $\mathbf{W}$ are learnable weights. It's a standard graph convolution layer that only considers neighbors from the positively or negatively linked neighborhood, $\mathcal{N}_i^+$ or $\mathcal{N}_i^-$. However, when ${l}$ > 1, the aggregation is more complex and can be defined as: 

\vspace{-5pt}
\begin{equation}
\mathbf{h}^{+(l)}_i = \sigma \left( \mathbf{W}^{+(l)} \left[ \sum_{j \in \mathcal{N}_i^+} \frac{\mathbf{h}^{+(l-1)}_j}{|\mathcal{N}_i^+|}, \sum_{k \in \mathcal{N}_i^-} \frac{\mathbf{h}^{-(l-1)}_k}{|\mathcal{N}_i^-|}, \mathbf{h}^{+(l-1)}_i \right] \right)
\end{equation}

\vspace{-5pt}
\begin{equation}
\mathbf{h}^{-(l)}_i = \sigma \left( \mathbf{W}^{-(l)} \left[ \sum_{j \in \mathcal{N}_i^+} \frac{\mathbf{h}^{-(l-1)}_j}{|\mathcal{N}_i^+|}, \sum_{k \in \mathcal{N}_i^-} \frac{\mathbf{h}^{+(l-1)}_k}{|\mathcal{N}_i^-|}, \mathbf{h}^{-(l-1)}_i \right] \right)
\end{equation}

For each node,  the positive embedding of a node comes from the positive set of embeddings from its positively linked neighborhood $\mathcal{N}_i^+$ (the friends of my friends); and the negative set of embeddings from its negatively linked neighborhood $\mathcal{N}_i^-$  (the enemies of my enemies); and the positive embedding of the node itself from last layer. Similarly, the negative embedding of a node comes from "the friends of my enemies", "the enemies of my friends" and the negative embedding itself from last layer. In the last layer, the embeddings of a node \(\mathbf{z}_i\) is the concatenation of the two sets of embeddings [\(\mathbf{h}^{+(l)}_i; \mathbf{h}^{-(l)}_i\)], and we further use Equation 2 to generate a new graph structure \(\mathcal{G}_{h}\), but in this section, we focus on whether the users and foods have enough similarity in the health aspect.

\subsubsection{Graph Structure Pooling}
Lastly, to reduce the noise in the real-life dataset, and enhance model's robustness, we further implement a structure pooling module to infuse the aforementioned two generated graph with our original graph structure and connect it to the downstream task in an end-to-end fashion, so that each graph structure is optimized over training. In this way, we aim at obtaining the best refined graph structure for our recommendation. Specifically, we fuse these graph adjacency matrix through a channel attention layer: 

\vspace{-5pt}
\begin{equation}
\mathcal{A}' = \textbf{Att} ([\mathcal{A}, \mathcal{A}_{ft}, \mathcal{A}_{sem}]),
\end{equation}
where \([\mathcal{A}, \mathcal{A}_{ft}, \mathcal{A}_{sem}]\) is the stacked matrix of the three graphs, and $\textbf{Att}$ denotes a self-attention layer with parameters which leverages their importance and perform a softmax function to obtain the final graph structure. Then, the refined structure is used as follows:  

\vspace{-5pt}
\begin{equation}
\mathbf{H}_l = \sum_{l=0}^{L} \left( \mathbf{D}^{-1/2} \mathcal{A}' \mathbf{D}^{-1/2} \right) \mathbf{H}_{(l-1)},
\end{equation}
where $L$ is the number of propagation layers. $\mathbf{D}$ is the degree matrix, with $D_{ii}$ representing the degree of node $i$. For simplicity, $\mathbf{H}^{(0)}$ is a randomly initialized embeddings. Through this way, MOPI-HFRS leverages both the diet preference patterns among users and the health and nutrition information for the recommendation and in the mean time, addresses the data noise with refined graph structure.

\subsection{Pareto Optimization}

Besides integrating information from multiple aspects into the training process, achieving the optimal recommendations that balance our multi-faceted objectives, requires the careful definition of a suitable objective function. The weighted sum approach, commonly used to transform multi-objective optimization problems into a single scalar objective in health-aware food recommendation, has significant drawbacks. One primary disadvantage is its inherent dependency on weight selection, which involves subjective judgments about the relative importance of each objective. Inappropriate weight selection may result in biased solutions that fail to reflect the true trade-offs among objectives. 


To address these limitations, we implement a multiple gradient descent algorithm that promotes Pareto optimality \cite{desideri2012multiple}. This algorithm aims to balance conflicting objectives, mitigate biases introduced by arbitrary weighting schemes, and provide a more robust and flexible framework for our multi-objective task. Specifically, we formulate our optimization problem as follows:

\vspace{-5pt}
\begin{equation}
\mathcal{L}_{\textit{Pareto}} = \min \left\| \sum_{k=1}^{K} \alpha_k \cdot \nabla_{\theta} \mathcal{L}_k (\mathcal{G}; \theta) \right\|_F,
\end{equation}
where $\nabla_{\theta} \mathcal{L}_k (\mathcal{G}; \theta)$ refers to the gradients of the model parameters for loss $k$. As shown in Figure-\ref{fig:framework}(d), optimizing multiple objectives involves finding the descent direction with the minimum norm within the convex hull defined by the gradient directions of each loss. Our task aims to integrate three potentially conflicting objectives: user preference, personalized healthiness, and nutrition diversity. For user preference, we use Bayesian Personalized Ranking (BPR) loss, which is widely used in recommendation tasks, defined as follows:

\vspace{-5pt}
\begin{equation}
\mathcal{L}_{BPR} = - \sum_{(u, i, j)} \ln \sigma(\hat{y}_{uij}) + \lambda \|\Theta\|^2,
\end{equation}
where $\sigma$ is the sigmoid function and $\hat{y}_{uij} = \hat{y}_{ui}$ - $\hat{y}_{uj}$ represents the difference in predicted scores between a preferred food $i$ and a non-preferred food $j$ for a user $u$, and $\sigma$ is the sigmoid function. $\lambda \|\Theta\|^2$ is a regularization term to prevent over-fitting. Following this, we designed two objective functions that reflects personalized healthiness and nutritional diversity of the recommendation, which can be formulated as follows: 

\vspace{-5pt}
\begin{equation}
\mathcal{L}_{\text{health}} = - \sum_{(u, i, j)} \ln \left( \left( \mathcal{J}(t_u, t_i) - \mathcal{J}(t_u, t_j) \right) \cdot \sigma(\hat{y}_{uij}) \right),
\end{equation}

\vspace{-7pt}
\begin{equation}
\mathcal{L}_{\text{diversity}} = - \sum_{(u, i, j)} \ln \sigma \left( \left(1 - \sum_{i, j \in \mathcal{N}_u} \cos (\mathbf{f}_i, \mathbf{f}_j) \right) \cdot \hat{y}_{ui} \right),
\end{equation}
where, for health loss, the $\mathcal{J}(t_u, t_i) - \mathcal{J}(t_u, t_j)$ measures the difference of the Jaccard similarity of the health tag vectors between a preferred food $i$ and a non-preferred food $j$ for a user $u$. For diversity loss, we calculate the average pair-wise cosine similarity between all predicted food embeddings of a user $\mathcal{N}_u$. Intuitively, the training process will pull the embeddings between users and predicted foods closer but for each user, the embeddings of predicted foods are pushed away from each other, thus minimizing this average similarity encourages diverse recommendations without directly hurting the recommendation performance. To obtain the Pareto optimal parameters, we utilize the ideas from Frank-Wolfe algorithms and prior works \cite{jaggi2013revisiting, ju2022multi}. Specifically, we initialize all $\alpha$ (in Equation 9) as 1/K, and at each iteration, we first identify the loss (denoted as t) whose descent direction correlates least with the current combined descent direction (i.e. \(\sum_{k=1}^{K} \alpha_k \cdot \nabla_{\theta} \mathcal{L}_k (\mathcal{G}; \theta)\), denoted as $\hat{\nabla}_{\theta}$ below), which can be found by: 

\vspace{-7pt}
\begin{equation}
t = \arg\min_i \hat{\nabla}_{\theta} \cdot \nabla_{\theta} \mathcal{L}_i(\mathcal{G}; \theta)^\top .
\end{equation}
We then iteratively update $\alpha$ using the following formula:

\vspace{-7pt}
\begin{equation}
\alpha := (1 - \eta) \cdot \alpha + \eta \cdot e_t, \quad \text{s.t.} \quad \eta = \frac{\hat{\nabla}_{\theta} \cdot \left( \hat{\nabla}_{\theta} - \nabla_{\theta} \mathcal{L}_t(\mathcal{G}; \theta) \right)^\top}{\left\| \hat{\nabla}_{\theta} - \nabla_{\theta} \mathcal{L}_t(\mathcal{G}; \theta) \right\|_F},
\end{equation}
where $\eta$ is the step size of the increment and $e_t$ is an one-hot vector with t-th element equal to one. This optimization will reconcile the multi-objective model we designed above and iteratively lead to Pareto optimization of our three objectives.

\begin{table*}[!t]
    \caption{Performance comparison between different models for food recommendation on different benchmarks. Results are reported as mean ± std\%. The best performance is bolded and runner-ups are underlined.}
    \vspace{-10pt}
    \resizebox{\linewidth}{!}{
        \begin{tabular}{l|ccccc|ccccc}
            \toprule
            \multirow{2}{*}{Model} & \multicolumn{5}{c|}{\textbf{Nutrition-macro only}} & \multicolumn{5}{c}{\textbf{Nutrition-all}} \vspace{2pt} \\
             & Recall@20 & NDCG@20 & H-Score@20 & AvgTags@20 & \% Foods@20 & Recall@20 & NDCG@20 & H-Score@20 & AvgTags@20 & \% Foods@20 \\ \midrule
            GCN\cite{kipf2016semi} & 10.19\small{±0.56} & 8.10\small{±0.54} & 35.44\small{±4.91} & 6.10\small{±0.09} & 0.653\small{±0.041} & 10.29\small{±0.60} & 8.07\small{±0.40} & 59.23\small{±6.60} & 10.83\small{±0.30} & 0.591\small{±0.066} \\
            
            GraphSAGE\cite{hamilton2017inductive} & 10.40\small{±0.61} & 8.28\small{±0.45} & 34.34\small{±5.04} & 6.09\small{±0.08} & 0.675\small{±0.078} & 10.50\small{±0.36} & 8.00\small{±0.42} & 58.74\small{±4.66} & 10.83\small{±0.23} & 0.628\small{±0.050} \\
            
            GAT\cite{velivckovic2017graph} & 10.61\small{±0.51} & 8.23\small{±0.46} & 37.34\small{±3.99} & \underline{6.12\small{±0.13}} & 0.601\small{±0.055} & 10.46\small{±0.56} & 8.25\small{±0.04} & 61.77\small{±6.97} & \underline{10.84\small{±0.19}} & 0.503\small{±0.058} \\ \midrule
            
            NGCF\cite{wang2019neural} & 11.62\small{±0.14} & 9.26\small{±0.10} & 36.72\small{±2.17} & 6.05\small{±0.49} & 0.678\small{±0.127} & 11.63\small{±0.15} & 9.16\small{±0.10} & 61.41\small{±1.63} & 10.83\small{±0.09} & 1.036\small{±0.403} \\
            
            LightGCN\cite{he2020lightgcn} & 11.68\small{±0.23} & 9.29\small{±0.13} & 35.83\small{±2.36} & 5.99\small{±0.05} & 0.567\small{±0.028} & \underline{11.80\small{±0.14}} & 9.23\small{±0.12} & 61.42\small{±1.52} & 10.81\small{±0.09} & 0.482\small{±0.024} \\
            
            SimGCL\cite{wu2021self} & \underline{11.73\small{±0.02}} & \underline{9.45\small{±0.01}} & \underline{37.59\small{±0.35}} & 6.04\small{±0.72} & 3.243\small{±0.621} & 11.74\small{±0.04} & \underline{9.32\small{±0.03}} & \underline{61.88\small{±0.74}} & 10.76\small{±0.02} & 2.889\small{±0.417} \\
            
            SGL\cite{yu2022graph} & 9.97\small{±0.17} & 7.67\small{±0.18} & 33.30\small{±1.24} & 6.01\small{±0.28} & \underline{4.459\small{±0.184}} & 9.31\small{±0.16} & 7.12\small{±0.14} & 56.42\small{±1.12} & 10.82\small{±0.06} & \underline{7.170\small{±0.452}} \\
            
            LightGCL\cite{cai2023lightgcl} & 10.97\small{±0.12} & 8.54\small{±0.11} & 35.77\small{±1.77} & 6.05\small{±0.04} & 3.597\small{±0.447} & 11.33\small{±0.10} & 8.70\small{±0.08} & 60.94\small{±1.48} & 10.82\small{±0.07} & 3.929\small{±0.448} \\ \midrule
            
            RecipeRec\cite{tian2022reciperec} & 10.03\small{±0.32} & 7.99\small{±0.41} & 35.94\small{±6.57} & 6.08\small{±0.16} & 2.070\small{±0.301} & 10.43\small{±0.51} & 8.02\small{±0.53} & 58.93\small{±8.01} & 10.77\small{±0.21} & 1.925\small{±0.280} \\
            
            HFRS-DA\cite{forouzandeh2024health} & 9.44\small{±0.62} & 7.34\small{±0.58} & 35.90\small{±0.53} & 6.04\small{±0.13} & 1.011\small{±0.072} & 9.59\small{±0.67} & 7.31\small{±0.63} & 59.87\small{±5.17} & 10.80\small{±0.26} & 0.979\small{±0.102} \\ \midrule
            
            \rowcolor{Gray} \textbf{MOPI-HFRS} & \textbf{12.91\small{±0.19}} & \textbf{10.34\small{±0.13}} & \textbf{38.13\small{±0.99}} & \textbf{6.17\small{±0.03}} & \textbf{10.550\small{±2.491}} & \textbf{12.22\small{±0.16}} & \textbf{9.64\small{±0.08}} & \textbf{63.07\small{±1.17}} & \textbf{10.94\small{±0.08}} & \textbf{16.510\small{±2.248}} \\
            \bottomrule
        \end{tabular}
    }
    \label{tab:results}
\end{table*}

\begin{table*}[!t]
    \vspace{-10pt}
    \caption{Performance comparison with different variants of our proposed MOPI-HFRS, including (I) Feature-based Structure Learning, (II) Healthy Edge Structure Learning, and (III) Pareto Optimization. Results are reported in the same format as above.}
    \vspace{-10pt}
    \resizebox{\linewidth}{!}{
        \begin{tabular}{ccc | ccccc | ccccc}
            \toprule
            & & & \multicolumn{5}{c|}{\textbf{Nutrition-macro only}} & \multicolumn{5}{c}{\textbf{Nutrition-all}} \vspace{2pt} \\
            \textbf{(I)} & \textbf{(II)} & \textbf{(III)} & Recall@20 & NDCG@20 & H-Score@20 & AvgTags@20 & \% Foods@20 & Recall@20 & NDCG@20 & H-Score@20 & AvgTags@20 & \% Foods@20 \\ \midrule
            $\surd$ & $-$ & $-$ & 12.69\small{±0.48} & 9.67\small{±0.49} & 36.13\small{±3.61} & 6.12\small{±0.12} & 0.483\small{±0.018} & 12.10\small{±0.47} & 9.14\small{±0.37} & 60.61\small{±3.47} & 10.91\small{±0.24} & 0.434\small{±0.011} \\
            
            $-$ & $\surd$ & $-$ & 12.76\small{±0.39} & 9.82\small{±0.41} & 35.36\small{±3.96} & \underline{6.14\small{±0.08}} & 0.492\small{±0.012} & 12.37\small{±0.35} & 9.16\small{±0.40} & 60.35\small{±2.89} & 10.93\small{±0.09} & 0.431\small{±0.016} \\
            
            $-$ & $-$ & $\surd$ & 11.47\small{±0.60} & 9.16\small{±0.44} & 36.67\small{±4.48} & 6.11\small{±0.08} & 1.551\small{±0.149} & 10.67\small{±0.54} & 8.35\small{±0.52} & 61.39\small{±5.11} & 10.88\small{±0.22} & 1.680\small{±0.132} \\
            
            $\surd$ & $-$ & $\surd$ & 12.85\small{±0.16} & 10.28\small{±0.10} & 37.35\small{±1.39} & 6.11\small{±0.05} & \underline{10.920\small{±2.302}} & 12.21\small{±0.11} & \underline{9.62\small{±0.06}} & 62.18\small{±1.79} & 10.89\small{±0.11} & \textbf{17.830\small{±2.953}} \\
            
            $-$ & $\surd$ & $\surd$ & 12.84\small{±0.18} & \underline{10.28\small{±0.08}} & \underline{37.63\small{±1.48}} & 6.13\small{±0.35} & \textbf{11.730\small{±1.241}} & 12.14\small{±0.10} & 9.59\small{±0.05} & \textbf{63.62\small{±0.83}} & \textbf{10.96\small{±0.07}} & 16.410\small{±2.58} \\
            
            $\surd$ & $\surd$ & $-$ & \textbf{13.02\small{±0.20}} & 10.19\small{±0.28} & 37.61\small{±4.67} & 6.12\small{±0.15} & 0.550\small{±0.037} & \textbf{12.46\small{±0.27}} & 9.36\small{±0.24} & 60.14\small{±3.27} & 10.90\small{±0.17} & 0.510\small{±0.088} \\ \midrule
            
            \rowcolor{Gray} $\surd$ & $\surd$ & $\surd$ & \underline{12.91\small{±0.19}} & \textbf{10.34\small{±0.13}} & \textbf{38.13\small{±0.99}} & \textbf{6.17\small{±0.03}} & 10.550\small{±2.491} & \underline{12.22\small{±0.16}} & \textbf{9.64\small{±0.08}} & \underline{63.07\small{±1.17}} & \underline{10.94\small{±0.08}} & \underline{16.510\small{±2.248}} \\
            \bottomrule
        \end{tabular}
    }
    \vspace{-10pt}
    \label{tab:ablation}
\end{table*}

\begin{figure}[!t]
  \centering
  \includegraphics[width=\linewidth]{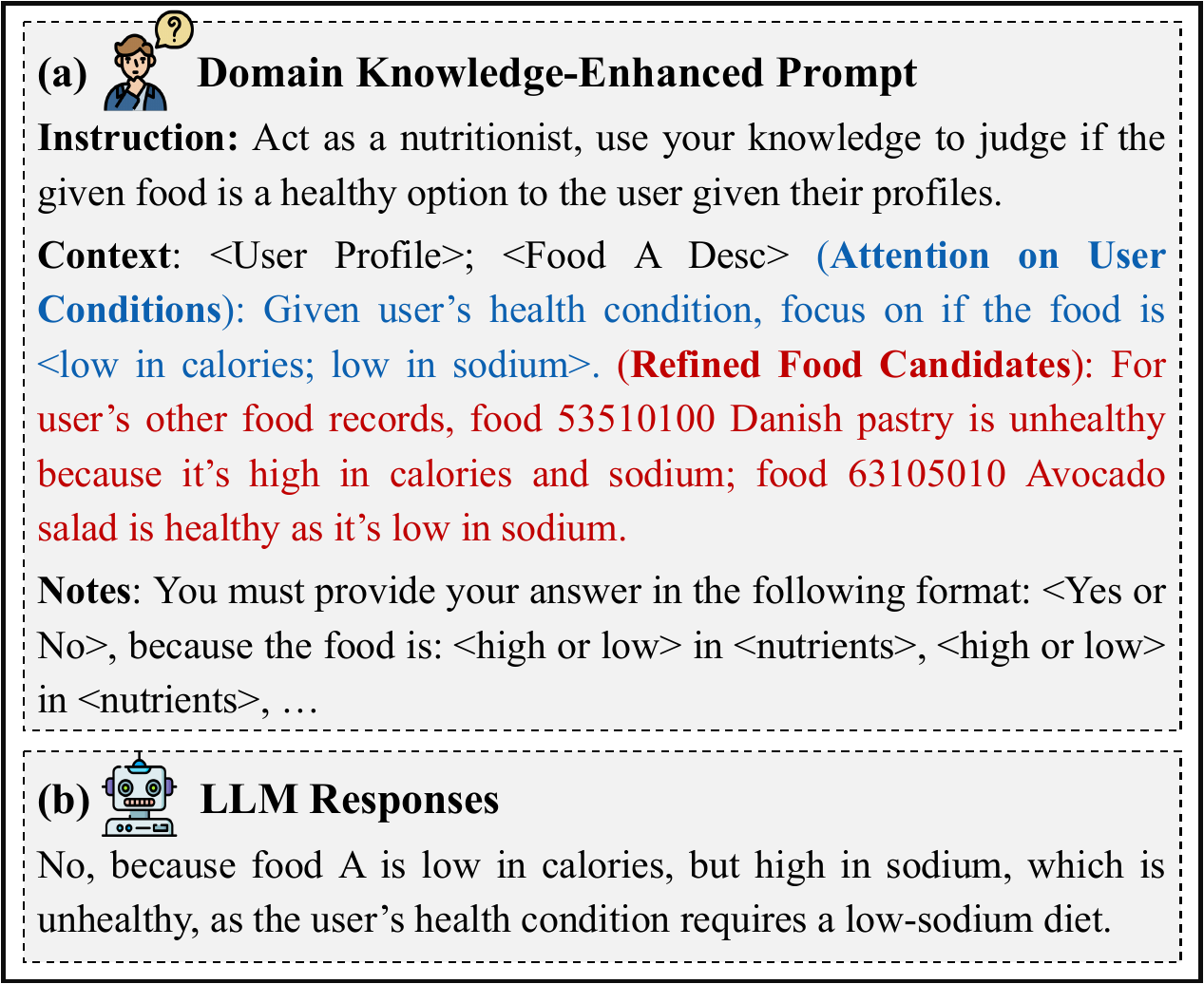}
  \vspace{-20pt}
  \caption{The prompts (highlighted in blue and red) generated from the MOPI-HFRS for interpreting the healthiness of recommendation results.}
  \label{fig:prompt}
  \vspace{-20pt}
\end{figure}

\subsection{Knowledge-infused LLM Reasoning}

Beyond recommending foods, our MOPI-HFRS framework also excels at interpreting the healthiness of these recommendations, which is crucial for improving users’ health awareness and guiding healthier dietary habits. To this end, our objective is to design prompts that enable LLM to perform trustworthy and reliable reasoning. As shown in Figure-\ref{fig:prompt}, a prompt consists of three parts: (1) Instruction, (2) Context, and (3) Notes. The Instruction provides task-specific directives, the Context offers tailored information, and the Notes guide LLM behavior, such as output length. Notably, the Context is the most crucial component, as it presents specific information. A simple approach to context design involves asking whether a particular food is healthy for a specific user, but this method does not fully leverage LLMs, especially in healthcare where domain-specific knowledge is vital \cite{tang2023does}. This can lead to reasoning biased by the LLM’s general knowledge, resulting in explanations that overlook critical health and nutrition insights. To address this, recent studies \cite{yang2023towards, blair2023can} have incorporated domain-specific knowledge into prompt generation, enriching reasoning capabilities. Inspired by these, we introduce two strategies to enhance LLM reasoning in our task, as shown in Figure-\ref{fig:prompt}(a).

\begin{figure}[!t]
  \centering
  \includegraphics[width=\linewidth]{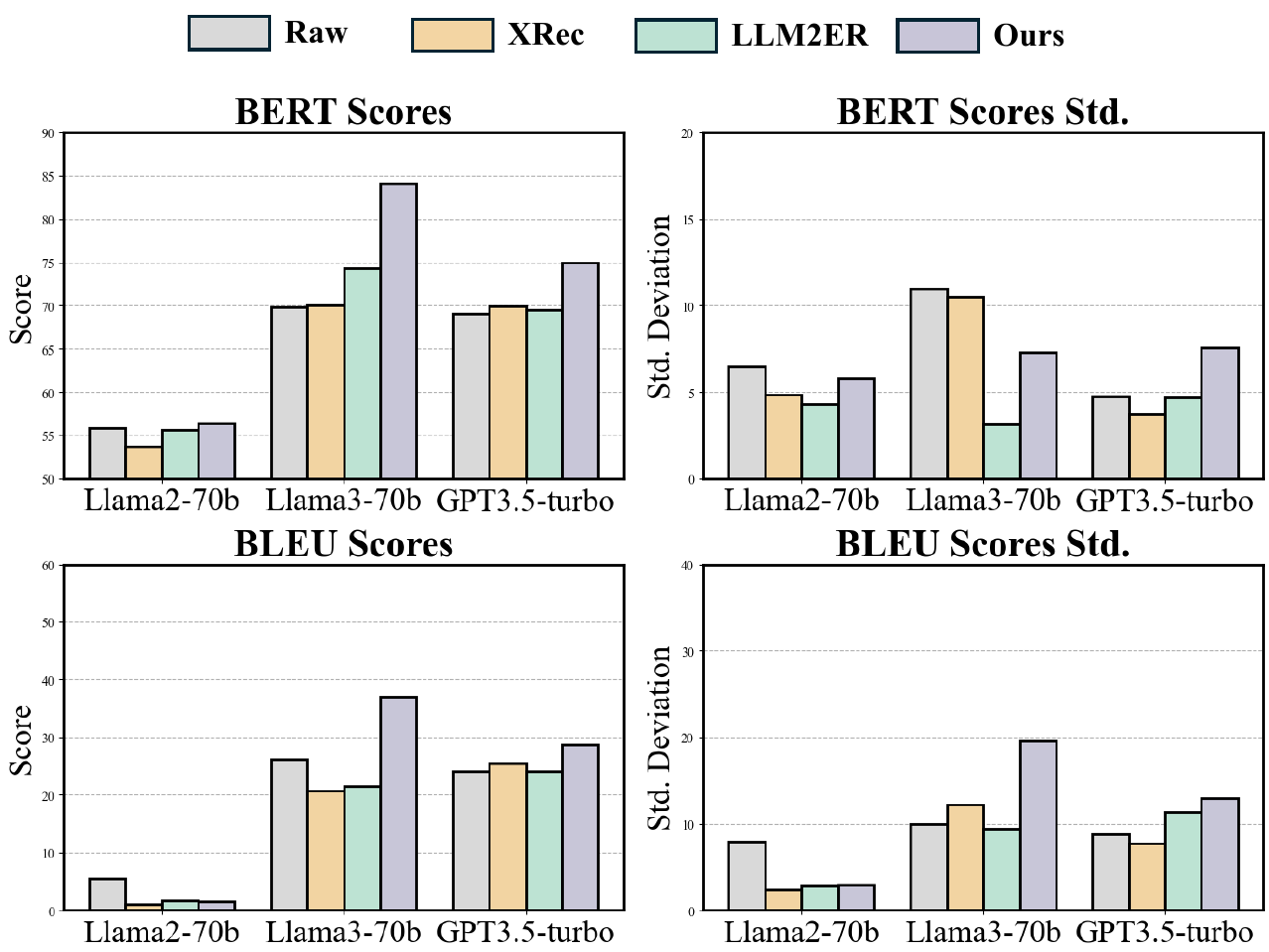}
  \vspace{-20pt}
  \caption{Performance of the reasoning task. Results are reported as mean and std. of BERT Scores and BLEU Scores.}
  \vspace{-15pt}
  \label{fig:rasoning}
\end{figure}

\textbf{\textit{Strategies for facilitating domain-specific reasoning.}}
We identify two primary limitations in current reasoning approaches. First, while examining what other foods are considered healthy for a user is beneficial, this approach falters when the user has interacted with many unhealthy foods or very few healthy ones, leading to conflicting information. Second, LLMs trained on general texts, may assume a food is healthy based on general knowledge, which may not suit users with specific health conditions. For example, a calorie-smart diet, typically seen as healthy, might not be suitable for users with low BMI, compromising reasoning reliability.

To address the first limitation, we introduce the \textit{Refined Food Candidates} strategy, which selects food candidates using the refined graph after training, reducing uncertainty from noisy data and optimizing for healthier options. To address the second, we implement the \textit{Attention on User Conditions} strategy, which emphasizes the health conditions most relevant to the user, helping LLMs capture the essential relationships between the user and recommended foods. Together, these strategies enable the LLM to provide more accurate and contextually appropriate analyses.

\section{Experiments}
\subsection{Experiment Setup}
\noindent\textbf{Baselines.} To demonstrate the superiority of our MOPI-HFRS in recommending diversified foods that both meets users' preference and health requirements, we select three major categories of graph-based baselines. These include 1) Classical graph neural network baselines: GCN~\cite{kipf2016semi}, GraphSAGE~\cite{hamilton2017inductive}, and GAT~\cite{velivckovic2017graph}, as they lay a great foundation of our analysis; 2) SOTA graph-based recommendation baselines: NGCF~\cite{wang2019neural}, LightGCN~\cite{he2020lightgcn}, SGL~\cite{wu2021self}, SimGCL~\cite{yu2022graph}, LightGCL~\cite{cai2023lightgcl}, as they are the golden standards of general recommendation systems; 3) SOTA food recommendation baselines: RecipeRec~\cite{tian2022reciperec}, HFRS-DA~\cite{forouzandeh2024health}, as they are the domain-specific SOTA models in the food recommendation field.

\begin{figure*}[htbp!]
	\centering
	\includegraphics[width=1\linewidth]{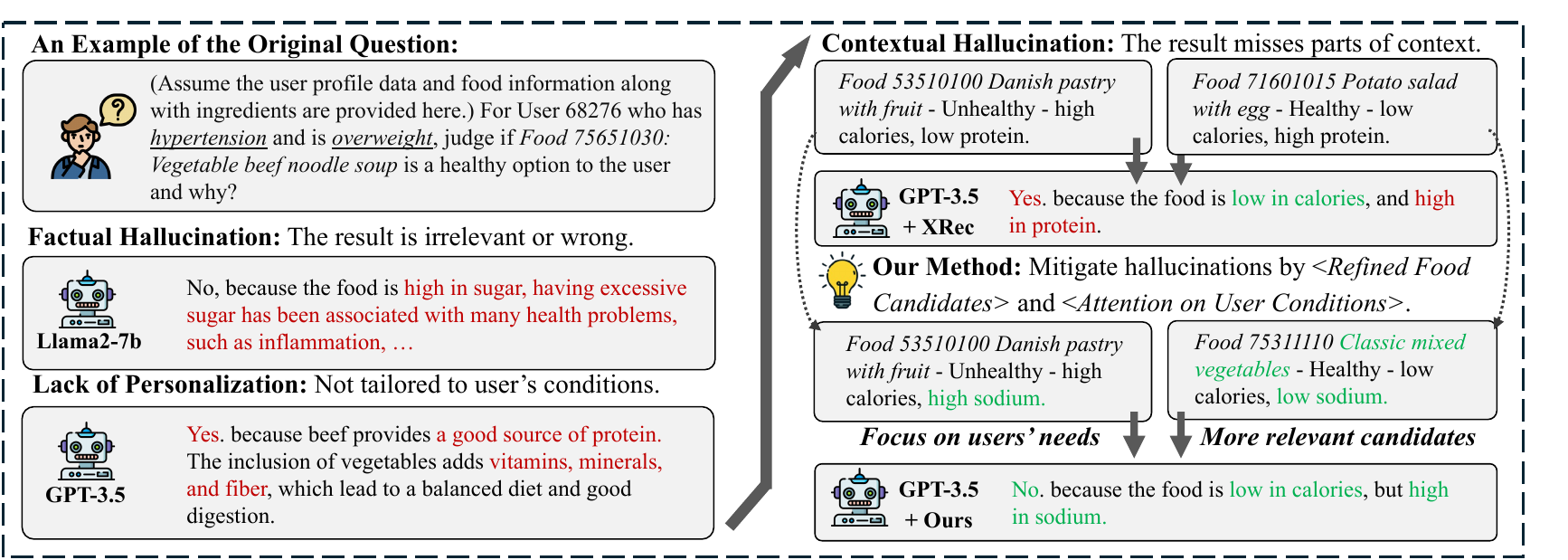}
        \vspace{-20pt}
	\caption{A case study demonstrating the three challenges faced by traditional LLM-enhanced interpretable methods. Our proposed MOPI-HFRS model effectively mitigates these issues and deliver tailored and accurate explanations.}
        \vspace{-15pt}
    \label{fig:case}
\end{figure*}

\noindent\textbf{Evaluation Settings.} In this study, we evaluate the two benchmarks using five metrics. Each model is run with ten consecutive seeds to mitigate the influence of randomness. To thoroughly assess the model's performance across our multiple objectives, we first employ two widely recognized metrics to evaluate the model performance on dietary preference: Recall and NDCG. To assess the healthiness of the recommendation, we develop a H-Score which extracts the health tags of a user and the predicted foods, and measures the average percentage of the predicted foods that share at least one common health tags with the user (i.e., the nutrients match the users' health requirements). To evaluate diversity, we follow the traditional scheme and develop AvgTags and Percentage (\%) of Foods to evaluate element-level and global-level diversity respectively. The AvgTags measures the number of different tags a user has been recommended in sum, so the model is encouraged to predict diversified foods that can cover various health requirements. The Percentage (\%) of Foods, on the other hand, measures the percentage of foods that have been recommended at least once, so the model is encouraged to recommend as many different foods as possible. More experiment settings about environment, split, hyper-parameters, and reasoning task setup are detailed in Appendix-A.

\vspace{-5pt}
\subsection{Results}
\noindent\textbf{Performance Comparison.} Table \ref{tab:results} demonstrates the performance of baseline models and our proposed MOPI-HFRS in the recommendation task. Notably, our MOPI-HFRS surpasses existing methods in all three aspects evaluated, while traditional methods exhibit limitations in balancing the trade-offs between the three tasks. As can be seen, the classical graph neural networks tend to have higher health scores and AvgTags, which might be attributed to their integration with features information, while current graph-based recommendation systems often solely utilize the structural information to improve efficiency and performance. As a result, these models tend to have higher recall and NDCG scores than the classical graph neural network baselines, whereas their healthiness scores are lower. Another notable pattern is that traditional methods suffer from the long tail problem \cite{park2008long, yin2012challenging}, which leads to recommending only the most popular foods, thus the significantly low percentage of the recommended foods. In contrast, our MOPI-HFRS addresses this challenge through its Pareto optimization process, thereby enhancing the model’s capability to balance multiple objectives.

\noindent\textbf{Ablation Study.} To identify the modules contributing to performance enhancement, we conduct a comprehensive ablation study, as shown in Table \ref{tab:ablation}. Notably, integrating only the Feature-based Structure Learning and Health-aware Structure Learning modules yields the highest recall, though performance in other areas, particularly the percentage of foods recommended, declines. This trend is evident when applying our Pareto loss to the raw LightGCN model, where health and diversity scores improve significantly with only a minor reduction in recall and NDCG, demonstrating the effectiveness of our Pareto Optimization module. Additionally, integrating Health-aware Structure Learning modules increases both health and Avgtag scores, affirming the value of incorporating health information during training. While each component independently enhances performance, the model achieves optimal results across all metrics when these three components are integrated, highlighting the synergistic effect of concurrently processing descriptive features, nutritional health information, and Pareto optimization.

\noindent\textbf{Reasoning analysis.}
Following the methodology design in Section 4.3, we employ the prompted texts for a generative reasoning task. Using the matched health tags as ground truth, we evaluate the quality of reasoning using BERT Score \cite{zhang2019bertscore} and BLEU Score \cite{post2018call}. To demonstrate MOPI-HFRS’s superiority, we compare it against two state-of-the-art LLM-enhanced explainable recommendation baselines—Xrec \cite{ma2024xrec} and LLM2ER \cite{yang2024fine}—using three popular LLM backbones: GPT-3.5-turbo, Llama-2-70b, and Llama-3-70b. Figure \ref{fig:rasoning} shows that MOPI-HFRS consistently outperforms other baselines in providing accurate personalized explanations, indicating the efficacy of our designed strategies. An exception is the BLEU score on Llama-2-70b, where lower scores may be due to the model’s difficulty with complex prompts, affecting all prompted baselines compared to the raw baseline. A detailed example of why our model outperforms others is explained in the following section. 

\vspace{-5pt}
\section{Case Study}
Our experimental results demonstrate the superiority of MOPI-HFRS in both multi-objective recommendation and reasoning tasks. To illustrate how our approach benefits downstream reasoning, we present a specific example in Figure \ref{fig:case}. As shown, traditional LLM-enhanced methods often struggle with domain-specific tasks like nutritional health, leading to issues such as factual hallucination, lack of personalization, and contextual hallucination.

In this example, we evaluate whether Vegetable Beef Noodle Soup is a healthy option for a user, who needs a low-calorie diet due to being overweight and a low-sodium diet due to hypertension. Factual hallucination occurs when LLMs like Llama-2-7b provide incorrect or irrelevant answers due to the complexity of the task, while powerful LLMs like GPT can produce correct results, they often lack personalization, overlooking user-specific health needs. 

This issue can be mitigated by prompt learning, as shown in the upper right of Figure \ref{fig:case}. By leveraging information from neighboring food nodes, LLMs can focus on key nutritional requirements, though contextual hallucination remains a challenge, where the model overlooks important information due to the overwhelming context. Our prompting strategies address this by using the two proposed strategies, emphasizing critical health aspects and identifying refined food candidates. Combining these strategies, MOPI-HFRS accurately determines that the food is high in sodium and unsuitable for the user, providing clear, personalized explanations that promote informed dietary decisions for user's future diet.

\section{Conclusion}

In this study, we presented MOPI-HFRS, a multi-objective LLM-enhanced interpretable personalized health-aware food recommendation system that integrates user dietary preferences with personalized health and nutrition diversity objectives. Extensive experiments using our built benchmarks have shown that MOPI-HFRS significantly outperforms state-of-the-art baselines in both recommendation performance and reliable explanations. The integration of graph-based learning with LLM-enhanced interpretability allows the system to deliver not only relevant food choices but also insightful interpretations, crucial for improving users’ understanding of healthy eating habits. We believe MOPI-HFRS opens new avenues for future research in enhancing the personalization and explainability of health-aware food recommendation systems.

\newpage
\bibliographystyle{ACM-Reference-Format}
\bibliography{reference}
\appendix
\section{Additional Experimental Settings}
\subsection{Environmental Settings}
The experiments are conducted in the Windows 10 operating system with 64GB of RAM. The training process involves the use of one NVIDIA GeForce RTX 3090 GPU and eight NVIDIA A40 GPU, with the framework of Python 3.8.18, Pytorch 2.1.0 and Pytorch-geometric 2.4.0.

\subsection{Splitting and Hyper-parameters}
We set k as 20 for simplicity. For a balanced comparison, all models are configured with a hidden dimension of 128, and utilize the Adam optimizer with a learning rate of 0.001 and a learning rate decay of every 200 epochs, with a fixed total number of epochs at 500. We perform Adaptive Moment Estimation (Adam) to optimize the models with a learning rate of 1e-3 and L2 regularization as 1e-6. If a baseline model fails to converge under these fixed settings, we tune it to achieve optimal performance. We randomly split the data into a 4-4-2 group (40\% training set, 40\% valid set and 20\% test set). The random split is only performed once and all experiments are done on the same split seed.  

\subsection{Reasoning Task Setup}
After the training, we randomly sample a subgraph of 200 records with 100 records being healthy for the user and the other unhealthy for the user. All reasoning tasks are done on the same sampled subgraph. We used openai API for GPT-3.5-turbo and llamaAI API for llama models. For baselines, we use the in-context learning part of the baselines for a fair comparison. 

\section{Data and Tagging Scheme}
NHANES Dataset is a publicly available and readily used datasets, with data releases every two years. In general, it has five sections of data: Demographics Data, Dietary Data, Examination Data, Laboratory Data and Questionnaire Data. In this paper, the nutrition information is gathered from the dietary sections of NHANES data. The food records in NHANES data are recorded using Food and Nutrient Database for Dietary Studies (FNDDS) food code. This database, integral to USDA, catalogs food and beverage consumption in What We Eat In America (WWEIA) database and NHANES data to help researchers conduct enhanced analysis of nutrient values in dietary intakes. By leveraging this dataset, we link the food items in NHANES dataset with food ingredients and WWEIA food category information to get food, ingredients, and category information. Then, using the provided nutritional data and the nutritional standards discussed in the paper, we are able to define the high and low thresholds for tagging foods, as shown in Table-\ref{tab:nutrient_thresholds}.

On the other hand, the user information comes from the demographic sections of NHANES data, as well as the laboratory, examination and questionnaire data.   The users' tags comes from two aspects: 1) if the users clearly states they are on a certain diet. 2) if a certain laboratory result of the users exceeds or below international health standards. Unlike nutrition thresholds, the thresholds of a certain laboratory result are well-defined and universal. For example, a BMI under 18.5 is underweight and over 25 is overweight. Here we list all the diets or laboratory results we check in this data: BMI; waist circumference; opioid misuse; blood urea nitrogen; low-density lipoprotein; blood pressure; red blood cell and osteoporosis as well as the 12 diets NHANES data provides, such as weight loss diets, diabetic diets, etc. We focus on the key laboratory results where nutritional diet can play an important rule. For example, it is well recommended that people with high blood pressure to consume less salt. 

With the above information, we create the two benchmarks, a simplified version of only considering macro nutrients and a full version, as described in the paper. Table-\ref{tab:nutrition_benchmarks} shows the statistics of the benchmarks.

\begin{table}[!t]
    \caption{Nutrient Reference Values (NRV) and thresholds (per 100g of food) used based on the nutritional standards}
    \vspace{-10pt}
    \centering
    \setlength{\tabcolsep}{10pt} 
    \resizebox{\linewidth}{!}{
        \begin{tabular}{lccc}
            \toprule
            \textbf{Nutrients} & \textbf{Low Threshold} & \textbf{High Threshold} & \textbf{NRV} \\
            \midrule\midrule
            Calories (kcal)      & 40   & 225   & 2000 \\
            Carbohydrates (g)    & 55   & 75    & -    \\
            Protein (g)          & 10   & 15    & 50   \\
            Saturated Fat (g)    & 1.5  & 5     & 20   \\
            Cholesterol (mg)     & 20   & 40    & 300  \\
            Sugar (g)            & 5    & 22.5  & -    \\
            Dietary Fiber (g)    & 3    & 6     & -    \\
            \midrule\midrule
            Sodium (mg)          & 120  & 200   & 2000 \\
            Potassium (mg)       & 0    & 525   & 3500 \\
            Phosphorus (mg)      & 0    & 105   & 700  \\
            Iron (mg)            & 0    & 3.3   & 22   \\
            Calcium (mg)         & 0    & 150   & 1000 \\
            Folic Acid (µg)      & 0    & 60    & 400  \\
            Vitamin C (mg)       & 0    & 15    & 100  \\
            Vitamin D (µg)       & 0    & 2.25  & 15   \\
            Vitamin B12 (µg)     & 0    & 0.36  & 2.4  \\
            \bottomrule
        \end{tabular}
    }
    \label{tab:nutrient_thresholds}
\end{table}

\begin{table}[!t]
    \caption{Statistics of the two benchmarks}
    \vspace{-10pt}
    \centering
    \setlength{\tabcolsep}{6pt} 
    \resizebox{\columnwidth}{!}{ 
        \begin{tabular}{lcccc}
            \toprule
            \textbf{Benchmarks} & \textbf{\# Users} & \textbf{\# Foods} & \textbf{\# Interactions} & \textbf{Sparsity} \\
            \midrule\midrule
            Nutrition\_all           & 13282 & 7516 & 488223 & 0.21\% \\
            Nutrition\_macro\_only    & 8170  & 6769 & 314224 & 0.22\% \\
            \bottomrule
        \end{tabular}
    }
    \label{tab:nutrition_benchmarks}
\end{table}

\section{Ethics and Privacy Statement}
Addressing privacy and ethical considerations is crucial, especially when handling sensitive health-related data. The National Health and Nutrition Examination Survey (NHANES) exemplifies best practices in this area, rigorously adhering to confidentiality safeguards as required by public legislation. This strong commitment to privacy allows us to achieve our core research objectives while remaining within the framework of established survey policies. Specifically, the original NHANES dataset has undergone an anonymization process to remove personally identifiable information (PII), such as social security numbers or physical addresses. Despite the absence of PII, the dataset remains a valuable resource for in-depth analysis, enabling us to explore the interaction between users’ medical data and health-aware food recommendations as discussed in this paper. Furthermore, in deployment, the usage of recommendation results and interpretations is safeguarded as part of personal medical records, ensuring continued privacy protection. By operating within these parameters, we ensure that our research upholds the highest standards of ethical integrity and privacy protection.

\section{Limitations and Discussion}
Although our developed \textit{MOPI-HFRS} has been demonstrated its outstanding performance on multi-objective recommendations and following reasoning task, it could also be subject to the limitation of lacking comparisons against other benchmarks. However this challenge is attributed to the sensitive and private nature of the topic. To our best knowledge, NHANES is the only public dataset that safely incorporates both user medical data and dietary information. There are no other benchmark datasets satisfying our requirements for the problem setting, and thus, we are unable to extend our experiments to other benchmarks under our problem setting. We hope this work can inspire future researchers and practitioners to work on this direction, and creating more benchmarks in the future. Another limitation is that in this work, we define diversity as various food items, whereas the definition of nutritional diversity is a far more delicate and complex topic. It's related to a careful combination of different categories of foods with different levels of nutrients and given the complexity of our model design, we leave this to our future work. Despite these limitations, our work in this paper is the first attempt to exploit personalized multi-objective health-aware food recommendations with explainable outputs, which could provide new insights and pave promising research directions for other researchers and practitioners toward this direction.

\end{document}